\documentclass{elsart}%
\usepackage{amsfonts}
\usepackage{amsmath}
\usepackage{amssymb}
\usepackage{natbib}
\usepackage{graphicx}
\begin{document}
\begin{frontmatter}
\title{Solid angle subtended by a cylindrical detector at a point source in terms of elliptic integrals}
\author{M. J. Prata\thanksref{FCT}}
\ead{mjprata@sapo.pt}
\thanks[FCT]{Partially supported by Funda\c{c}\~{a}o para a Ci\^{e}ncia e Tecnologia
(Programa Praxis XXI - BD/15808/98)}
\address{Instituto  Tecnol\'{o}gico e Nuclear (ITN), Sacav\'{e}m, Portugal}
\begin{abstract}
The solid angle subtended by a right circular cylinder at a point source located at an arbitrary position
generally consists of a sum of two terms: that defined by the cylindrical surface ($\Omega _{cyl}$) and the other by either
of the end circles ($\Omega _{circ}$).
We derive an expression for $\Omega _{cyl}$ in terms of elliptic integrals of the first
and third kinds and give similar expressions for $\Omega _{circ}$ using integrals
of the first and second kinds. These latter can be used alternatively to an expression
also in terms of elliptic integrals, due to Philip A. Macklin and
included as a footnote in Masket (Rev. Sci. Instr., 28 (3), 191-197, 1957).
The solid angle subtended by the whole cylinder when the source is
located at an arbitrary location can then be calculated using elliptic integrals.
\end{abstract}
\end{frontmatter}

\section{Introduction}

The knowledge of the solid angle ($\Omega$) subtended by a right, finite,
circular cylinder at a point isotropic source is required in numerous problems
in nuclear and radiation physics. Generally, $\Omega$ can be expressed as sum
of two components: that subtended by the cylindrical surface ($\Omega_{cyl}$)
and the other by either of the end circles ($\Omega_{circ}$). Through the
years this calculation has been addressed by various authors using different
methods. Without the worry of being exhaustive we give some examples of such
works. \citet{Mask57} outlined a general procedure based on Stokes theorem to
reduce the double integral $\Omega={\textstyle\iint}\sin\theta d\theta d\varphi$ 
to a contour integral in a single variable ($\theta$ or $\varphi$). 
The method was used to express $\Omega_{circ}$ and
$\Omega_{cyl}$ as single integrals which were numerically integrated.
Extensive tables describing these results both for the disc and the entire
cylinder were reported in a separate work \citep{Mask56}. An approximation to
the solid angle defined by two parallel plane surfaces was described by
\citet{Gill70} and applied in the cases of two equal rectangles and two equal
circles in a face-to-face geometry.\ With this method each surface is
subdivided into small finite areas and the two double integrals are then
replaced by a double summation. The calculation of $\Omega_{circ}$ was treated
by \citet{Gard71} by replacing the disc with a regular n-side polygon of equal
area, for which an analytical expression was given. In a similar way,
$\Omega_{cyl}$ was approximated by the analytical expression for an n-side
regular polyhedral surface \citep{Verg72}. \citet{Gree74} used the Monte Carlo
method to calculate the cylinder solid angle for two height-to-radius ratios
(1:1, 2:1), considering distances from source to cylinder center up to 12
cylinder radii and angular positions of the source ranging from $0{{}^o}$
to $90{{}^o}$ from the cylinder axis.

An analytical expression for $\Omega_{circ}$ in terms of elliptic integrals
due to Philip A. Macklin \citep{Mack57} appears included as a footnote in 
\citet{Mask57}. In the present work we show that also $\Omega_{cyl}$ can be reduced to
elliptic integrals and give, without derivation, expressions for
$\Omega_{circ}$ which can be deduced in a akin way and are different from that
due to P.A. Macklin. The solid angle of the whole cylinder can then be
expressed in terms of elliptic integrals which are rather well known functions
\citep[e.g.][]{Miln64} for which computation algorithms and tables are readily available.

Instead of blind numerical integration, one can turn to the variety of numeric methods 
already existent, which enable the fast calculation of the solid angle for the whole range
of parameters. For instance, the complete integrals of the first and second kinds
can be computed using (i) the polynomial approximations due to \citet{Hast55} and
included in \citet[][eqs. 17.3.33 to 17.3.36]{Miln64}; (ii) the process of the 
arithmetic-geometric mean \citep[][17.6]{Miln64} or (iii) the infinite series 
\citep[][17.3.11, 17.3.12]{Miln64} which can be used in combination with
Landen's transformation when the modular angle is close to $\pi /2$. 

Since the solid angle can be decomposed into elliptic integrals, any possibility
of finding general analytical expressions in terms of elementary functions is 
henceforward precluded. On the other hand, the calculation has been put under the 
sound roof of the subject of elliptic integrals and functions.

In a recent work \citep{Prata2002} we describe the analytical calculation of
the solid angle subtended by a cylinder at a point cosine source. Combining
these results with those presented here, the case of an axially symmetric point
source with an angular distribution given by $f_{k}(\mathbf{\Omega
})=1+a~\mathbf{\Omega\cdot k}$ can be treated analitically when the source
axis ($\mathbf{k}$) is orthogonal to that of the cylinder.

\section{Solid Angle Calculation\label{section_solid_angle}}

\begin{figure}[h]
\begin{center}
\includegraphics[height=4.7579cm,width=5.0918cm]{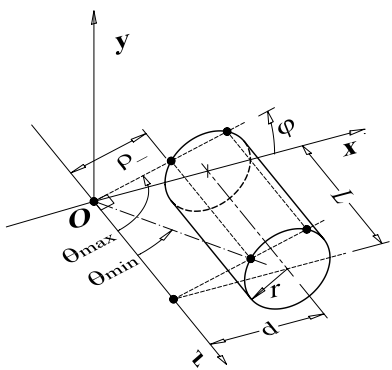}
\end{center}
\caption{Notation for $\Omega_{cyl0}$}%
\label{fig1}%
\end{figure}

\begin{figure}[h]
\begin{center}
\includegraphics[
height=4.7579cm,
width=5.0918cm
]{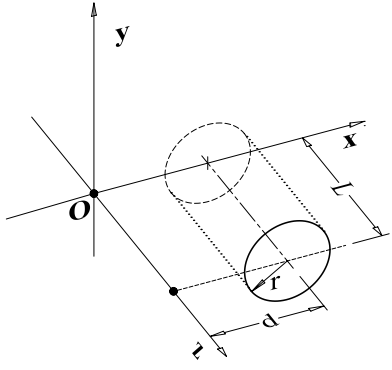}
\end{center}
\caption{Notation for $\Omega_{circ}$}%
\label{fig2}%
\end{figure}

The solid angle subtended by a given surface at a point isotropic source can
be defined as%

\begin{equation}
\Omega_{surf}=\frac{1}{4\pi}\iint
\limits_{\substack{directions\\hitting~surface}}d\Omega~,
\end{equation}
so that $0\leq\Omega_{surf}\leq1$. As previously said, in the case of a
circular cylinder, the solid angle $\Omega$ is in general given by
$\Omega=\Omega_{cyl}+\Omega_{circ}$, where $\Omega_{cyl}$ and $\Omega_{circ}$
are the contributions of the cylindrical surface and of one of the end
circles. These quantities can be defined with the help of figs. \ref{fig1} and
\ref{fig2}, where the source is located at the origin of the coordinate system
and the $z$ axis is both parallel to the cylinder axis and orthogonal to the
planes of the circles. To calculate $\Omega_{cyl}$ it is sufficient to
consider the situation depicted in fig. \ref{fig1}, where the source lies in
one of the planes delimiting the cylindrical surface. Let $\Omega
_{cyl0}(L,r,d)$ denote the solid angle in this case and $\Omega_{circ}(L,r,d)$
the solid angle defined by the disc. Let us illustrate how $\Omega$ can be
calculated in an arbitrary situation. Referring again to fig. \ref{fig1}, but
assuming \ that the source is located at $(0,0,-\left\vert z\right\vert )$,
one has $\Omega=\Omega_{cyl0}(L+\left\vert z\right\vert ,r,d)-\Omega
_{cyl0}(\left\vert z\right\vert ,r,d)+\Omega_{circ}(\left\vert z\right\vert
,r,d)$. If the source is located at $(0,0,z)$ and $0<z<L$, then $\Omega
=\Omega_{cyl0}(z,r,d)+\Omega_{cyl0}(L-z,r,d)$. If $r>d$ and $z<0$ or $z>L$,
the solid angle is defined by one of the circles alone, e.g. $\Omega
=\Omega_{circ}(\left\vert z\right\vert ,r,d)$ for $z<0$.

\subsection{Calculation of $\Omega_{cyl0}$}

\begin{figure}[h]
\begin{center}
\includegraphics[
height=3.6091cm,
width=5.0918cm
]{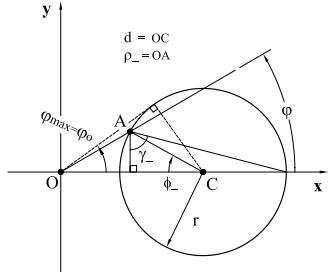}
\end{center}
\caption{{}Quantities used to calculate $\Omega_{cyl0}$}%
\label{fig3}%
\end{figure}From figs. \ref{fig1} and \ref{fig3}, it follows that

$\Omega_{cyl0}(L,r,d)=1/(4\pi)\int\limits_{\varphi_{\min}}^{\varphi_{\max}%
}\int\limits_{\theta_{\min}}^{\theta_{\max}}\sin\theta~d\theta d\varphi
=1/(2\pi)\int\limits_{0}^{\varphi_{o}}(\cos\theta_{\min}-\cos\theta_{\max
})~d\varphi$

where%

\begin{equation}
\varphi_{\max}=-\varphi_{\min}=\varphi_{o}\equiv\arcsin(r/d)~,
\label{eq_phio_def}%
\end{equation}

$\cos\theta_{\min}=L/\sqrt{L^{2}+\rho_{-}^{2}(\varphi)}~,$

$\cos\theta_{\max}=0$

and%

\begin{equation}
\rho_{-}(\varphi)=d\cos\varphi-\sqrt{r^{2}-\left(  d\sin\varphi\right)  ^{2}%
}~. \label{eq_rho_minus_def}%
\end{equation}

Thus,%

\begin{equation}
\Omega_{cyl0}=L/(2\pi)\int\limits_{0}^{\varphi_{o}}\left[  L^{2}+\rho_{-}%
^{2}(\varphi)\right]  ^{-1/2}~d\varphi~. \label{eq_omega_cylo_integral_phi}%
\end{equation}

Then we change the integration variable to $\gamma_{-}$ represented in fig.
\ref{fig3} and given by%

\begin{equation}
\gamma_{-}=\pi/2-\phi_{-}/2 \label{eq_gamma_phi}%
\end{equation}

where

$\phi_{-}/2=\arctan\left[  \sin(\varphi)\rho_{-}/(r+d-\cos(\varphi)\rho
_{-})\right]  $

and $\rho_{-}(\varphi)$ is obtained from eq. \ref{eq_rho_minus_def}.

Eq. \ref{eq_omega_cylo_integral_phi} is rewritten as%

\begin{equation}
\Omega_{cyl0}=L/(2\pi)\int\limits_{\gamma_{o}}^{\pi/2}\frac{1}{\sqrt
{L^{2}+\rho_{-}^{2}(\gamma_{-})}}\left[  \frac{d^{2}-r^{2}}{\rho_{-}%
^{2}(\gamma_{-})}-1\right]  ~d\gamma_{-}~ \label{eq_omega_cylo_integral_gamma}%
\end{equation}

where%

\begin{equation}
\rho_{-}(\gamma_{-})=\sqrt{(d+r)^{2}-4dr\sin^{2}\gamma_{-}}
\label{eq_rho_minus_gamma}%
\end{equation}

and%

\begin{equation}
\gamma_{o}=(\pi/2+\varphi_{o})/2~. \label{eq_gamma_o_def}%
\end{equation}

Introducing%

\begin{equation}
m=4rd/(L^{2}+\left(  d+r\right)  ^{2})~, \label{eq_m_def}%
\end{equation}

\begin{equation}
n=4rd/\left(  d+r\right)  ^{2}~, \label{eq_n_def}%
\end{equation}

there results from eq. \ref{eq_rho_minus_gamma}:

$\rho_{-}^{2}(\gamma_{-})=\left(  d+r\right)  ^{2}(1-n\sin^{2}\gamma_{-})~,$

$\sqrt{L^{2}+\rho_{-}^{2}(\gamma_{-})}=\sqrt{L^{2}+\left(  d+r\right)  ^{2}%
}\sqrt{1-m\sin^{2}\gamma_{-}}~,$

$L/\sqrt{L^{2}+\left(  d+r\right)  ^{2}}=\sqrt{1-m/n}~$

and

$(d-r)/(d+r)=\sqrt{1-n}~;d\geq r~.$

Substituting in the rhs of eq. \ref{eq_omega_cylo_integral_gamma} yields

$\Omega_{cyl0}=1/(2\pi)\sqrt{1-m/n}\int\limits_{\gamma_{o}}^{\pi/2}\frac
{1}{\sqrt{1-m\sin^{2}\gamma_{-}}}\left[  \frac{\sqrt{1-n}}{(1-n\sin^{2}%
\gamma_{-})}-1\right]  ~d\gamma_{-}~.$

The integral is easily decomposed into elliptic integrals in the Legendre form:%

\begin{equation}
\Omega_{cyl0}=1/(2\pi)\sqrt{1-m/n}\{\sqrt{1-n}[\pi(n;m)-\pi(n;\gamma
_{o}|m)]-[K(m)-F(\gamma_{o}|m)]\} \label{eq_omega_cylo_integral_elliptical}%
\end{equation}

where $F(\gamma_{o}|m)$, $\pi(n;\gamma_{o}|m)$ are the incomplete elliptic
integrals of the first and third kinds, respectively; and $K(m)=F(\pi/2|m)$,
$\pi(n;m)=\pi(n;\pi/2|m)$ are the corresponding complete integrals
\citep{Miln64}.\ From eqs. \ref{eq_m_def} and \ref{eq_n_def} it is clear that
the \textit{parameter} $m$ and the \textit{characteristic} $n$ verify%

\begin{equation}
0\leq m\leq1~,(m=1\Leftrightarrow L=0\wedge d=r)~, \label{eq_m_less_1}%
\end{equation}

\begin{equation}
0\leq n\leq1~,(n=1\Leftrightarrow d=r) \label{eq_n_less_1}%
\end{equation}

and%

\begin{equation}
0\leq m\leq n\leq1~;(m=n\Leftrightarrow L=0)~. \label{eq_m_less_n}%
\end{equation}

For computational purposes it is worth mentioning that, because of eq.
\ref{eq_m_less_n}, both $\pi(n;\gamma_{o}|m)$ and $\pi(n;m)$ in the rhs of eq.
\ref{eq_omega_cylo_integral_elliptical} fall into the the circular case
\citep[][17.7.11]{Miln64}, except when $L=0$ or $d=r$. These special cases are
treated separately in section \ref{section_special_values}.

\subsubsection{$\Omega_{cyl0}$ when $d\gtrapprox r$}

From eqs. \ref{eq_gamma_o_def} and \ref{eq_phio_def} results that, as
$d\rightarrow r^{+}$, $\gamma_{o}\rightarrow\pi/2$ and the two subtractions of
elliptic integrals in the rhs of eq. \ref{eq_omega_cylo_integral_elliptical}
approach zero (e.g. $K(m)-F(\gamma_{o}|m)\rightarrow0$). To avoid the
consequent loss of precision, an approximation\footnote{We include here a
normalizing factor of $1/(4\pi)$.} \citep[][eq. 32]{Mask57} can be used:%

\begin{align}
\Omega_{cyl0}  &  =(2\pi)^{-1}\{\varphi_{o}-1/2[rd\cos\varphi_{o}-r^{2}%
(\pi/2-\varphi_{o})]L^{-2}\nonumber\\
&  +3/8[rd(d^{2}+2r^{2})\cos\varphi_{o}-r^{2}(r^{2}+2d^{2})(\pi/2-\varphi
_{o})]L^{-4}-\ldots\}~, \label{eq_omega_cylo_series}%
\end{align}
where the cancellations can be more easily dealt with. Eq.
\ref{eq_omega_cylo_series} results from expanding and termwise integrating the
integrand on the rhs of eq. \ref{eq_omega_cylo_integral_phi}. Since
$d-r\leq\rho_{-}(\varphi)\leq\sqrt{d^{2}-r^{2}}$ when $0\leq\varphi\leq
\varphi_{o}$, the expansion can safely be done for $\sqrt{d^{2}-r^{2}}<L$ and
eq. \ref{eq_omega_cylo_series} gives a good approximation (better than four
digits) for $\sqrt{d^{2}-r^{2}}<L/10$.

\subsection{Expressions for $\Omega_{circ}$}

In a similar fashion to that used to obtain $\Omega_{cyl0}$ one can show that
the solid angle subtended by a circular disc (fig. \ref{fig2}) can be written
using complete elliptic integrals of the third and first kinds:%

\begin{equation}
\Omega_{circ}(L,r,d)=(2\pi)^{-1}\sqrt{1-m/n}\{\sqrt{1-n}~\pi(n;\pi
/2|m)-K(m)\}~;d>r \label{eq_omega_circ_integral_elliptical_dbigger}%
\end{equation}

\begin{equation}
\Omega_{circ}(L,r,d)=1/2-(2\pi)^{-1}\sqrt{1-m/n}\{\sqrt{1-n}~\pi
(n;\pi/2|m)+K(m)\}~;d<r \label{eq_omega_circ_integral_elliptical_rbigger}%
\end{equation}

where $m$, $n$ are obtained from eqs. \ref{eq_m_def}, \ref{eq_n_def}.

The complete integral of the third kind can be expressed in terms of
incomplete integrals of the first and second kinds which are easier to
calculate. Since $0\leq m\leq n\leq1$ one can use
\citep[][17.7.14,17.4.40]{Miln64} to show that:%

\begin{gather}
\Omega_{circ}=1/4-(2\pi)^{-1}n/(1+\sqrt{1-n})\sqrt{1-m/n}K(m)\nonumber\\
-(2\pi)^{-1}\{[E(m)-K(m)]F(\epsilon|m^{\prime})+K(m)E(\epsilon|m^{\prime
})\}~,d>r~, \label{eq_omega_circ_elliptical12_dbigger}%
\end{gather}

\begin{gather}
\Omega_{circ}=1/4-(2\pi)^{-1}(1+\sqrt{1-n})\sqrt{1-m/n}K(m)\nonumber\\
+(2\pi)^{-1}\{[E(m)-K(m)]F(\epsilon|m^{\prime})+K(m)E(\epsilon|m^{\prime
})\}~,d<r~, \label{eq_omega_circ_elliptical12_rbigger}%
\end{gather}

where%

\begin{equation}
m^{\prime}=1-m~, \label{eq_m_complementary_def}%
\end{equation}

\begin{equation}
\epsilon=\arcsin\sqrt{(1-n)/(1-m)}~,
\end{equation}

$E(\varphi|m)$ is the elliptic integral of the second kind and $E(m)=E(\pi
/2|m)$ is the complete integral.

As an alternative to eqs. \ref{eq_omega_circ_elliptical12_dbigger},
\ref{eq_omega_circ_elliptical12_rbigger}, one can use an
expression\footnote{We introduce a factor of $4\pi$ to conform to the
normalization used in the present work.} due to \citet{Mack57} which appears as
a footnote in \citet{Mask57}:%

\begin{gather}
(4\pi)\Omega_{circ}(L,r,d)=2\pi+2[K(m)-E(m)][F(\theta|m^{\prime}%
)+F(\psi|m^{\prime})]\nonumber\\
-2K(m)\left\{  E(\theta|m^{\prime})+E(\psi|m^{\prime})+2\beta/([1+(\alpha
+\beta)^{2}]^{\frac{1}{2}}[\beta+(1+\alpha^{2})^{\frac{1}{2}}])\right\}  ~,
\label{eq_omega_circ_elliptical_macklin}%
\end{gather}

where

$\theta=\arcsin\frac{[1+(\alpha+\beta)^{2}]^{\frac{1}{2}}}{\beta+(1+\alpha
^{2})^{\frac{1}{2}}}~,$ $\ \ \psi=\arcsin\frac{(1+\alpha^{2})^{\frac{1}{2}%
}-\beta}{[1+(\alpha-\beta)^{2}]^{\frac{1}{2}}}~,$

$0\leq\theta\leq\pi/2~,$ $\ \ -\pi/2\leq\psi\leq\pi/2~,$ $\ \ \theta
\geq\left\vert \psi\right\vert ~,$

$0\leq\alpha=d/L\leq\infty~,$ $\ \ 0\leq\beta=r/L\leq\infty$

and $m$, $m\prime$ are obtained from eqs. \ref{eq_m_def} and
\ref{eq_m_complementary_def}.

\subsection{Special values and continuity\label{section_special_values}}

Since $K(m=1)=\infty$ and $\pi(n;\pi/2|m)=\infty$ whenever $n=1$ or $m=1$, the
cases $L=0$ and $d=r$ have to be studied separately (see eqs.
\ref{eq_m_less_1} and \ref{eq_n_less_1}). Starting from eq.
\ref{eq_omega_cylo_integral_phi} it is straightforward that $\Omega
_{cyl0}(d\longrightarrow r^{+},L\neq0)=1/4$. Since $\Omega_{cyl0}(d>r,L=0)=0$,
$\Omega_{cyl0}$ is discontinuous when $d\longrightarrow r^{+}$
,$L\longrightarrow0$. Using the properties of elliptic integrals
\citep{Miln64} one can easily show from eqs.
\ref{eq_omega_circ_elliptical12_dbigger} and
\ref{eq_omega_circ_elliptical12_rbigger} that:%

\begin{equation}
\Omega_{circ}(d\longrightarrow r^{\pm},L\neq0)=1/4-(2\pi)^{-1}\sqrt{1-m_{1}%
}~K(m_{1})~, \label{eq_omega_circ_reqd}%
\end{equation}

$\Omega_{circ}(L\longrightarrow0)=\left\{
0~(d>r);1/4~(d=r);1/2~(d<r)\right\}  $

where $m_{1}=m|_{r=d}=4r^{2}/(L^{2}+4r^{2})$. $\Omega_{circ}$ is thus
continuous except for $L=0$. Eq. \ref{eq_omega_circ_reqd} is the same as
\citet[][eq. 24]{Mask57}, divided by a normalizing factor of $4\pi$.

\section{Summary and outlook}

An expression for $\Omega_{cyl}$ in terms of complete and incomplete elliptic
integrals of the first and third kinds has been derived in the case where the
source lies in the same plane as one of the end discs (eq.
\ref{eq_omega_cylo_integral_elliptical} ). Expressions for $\Omega_{circ}$
decomposed into complete and incomplete elliptic integrals of the first and
second kinds using a single amplitude ($\epsilon$) were given (eqs.
\ref{eq_omega_circ_elliptical12_dbigger} and
\ref{eq_omega_circ_elliptical12_rbigger}). These expressions can be used
alternatively to one (eq.\ref{eq_omega_circ_elliptical_macklin}) due to 
\citet{Mack57} containing also complete and
incomplete elliptic integrals of the first and second kinds but using two
amplitudes ($\theta,\psi$). Computational methods for the elliptic integrals
can be found in \citet{Miln64}. The required third kind integrals all belong
to the circular case ($0\leq m<n<1$) except for a few cases which are treated
in section \ref{section_special_values}. The solid angle when the source is at
an arbitrary position can be calculated using the expressions referred, as
described in the beginning of section \ref{section_solid_angle}.

The solid angle defined by a right circular cylinder and an isotropic source
distributed on a wire parallel to the cylinder axis can also be decomposed
into elliptic integrals in a way akin to that used here to obtain
$\Omega_{cyl0}$. A work where we report this result is in preparation.

\begin{ack}
Thanks are due to Jo\~{a}o Prata for reviewing this manuscript.
I would like to thank Professor John H. Hubbell for
providing a copy of the work by A.V. Masket \citep{Mask57}.
\end{ack}

\end{document}